\documentclass[12pt]{iopart}

\usepackage{cite}

\begin{document}

\title[PT-symmetric Hamiltonians]{Comment on: `Numerical estimates of the spectrum for anharmonic PT symmetric
potentials' [Phys. Scr. \textbf{85} (2012) 065005]}
\author{Paolo Amore\dag \ and Francisco M Fern\'andez
\footnote[2]{Corresponding author}}

\address{\dag\ Facultad de Ciencias, Universidad de Colima, Bernal D\'iaz del
Castillo 340, Colima, Colima, Mexico}\ead{paolo.amore@gmail.com}

\address{\ddag\ INIFTA (UNLP, CCT La Plata-CONICET), Divisi\'on Qu\'imica Te\'orica,
Blvd. 113 S/N,  Sucursal 4, Casilla de Correo 16, 1900 La Plata,
Argentina}\ead{fernande@quimica.unlp.edu.ar}

\maketitle

\begin{abstract}
We show that the authors of the commented paper draw their
conclusions from the eigenvalues of truncated Hamiltonian matrices
that do not converge as the matrix dimension increases. In one of
the studied examples the authors missed the real positive
eigenvalues that already converge towards the exact eigenvalues of
the non-Hermitian operator and focused their attention on the
complex ones that do not. We also show that the authors misread
Bender's argument about the eigenvalues of the harmonic oscillator
with boundary conditions in the complex-$x$ plane(Rep. Prog. Phys.
{\bf 70} (2007) 947.
\end{abstract}

\section{Introduction}

In a recent paper Bowen et al\cite{BMFM12} discussed the spectra
of a class of non-Hermitian hamiltonians having PT symmetry. They
calculated the eigenvalues of truncated matrices for the
non-Hermitian Hamiltonian operators in the basis set of the
eigenfunctions of the harmonic oscillator and argued that their
results did not agree with those of Bender and
Boettcher\cite{BB98}. They concluded that the discrepancy may be
due to the fact that the WKB method used by the latter authors is
unsuitable for such problems. For example, they stated that ``It
is certainly not obvious that the physical conditions of the
Bohr-Sommerfeld procedure should be valid for this non-physical
path. The significant differences in the spectrum studied in this
paper suggests that it is not valid'' and also that ``It is not
clear whether the motion along paths in the complex plane has any
physical significance for quantization''. Curiously, the autors
did not appear to pay attention to other methods for the
calculation of the eigenvalues of those PT-symmetric Hamiltonians.
For example, the WKB results were confirmed by numerical
integration based on the Runge-Kutta algorithm\cite{BB98,B07} as
well as by diagonalization of a truncated Hamiltonian matrix in
the basis set of harmonic-oscillator eigenfunctions\cite{BB98} (a
more detailed description of this approach was given in an earlier
version of the published paper\cite{BB97}). In addition to it,
Handy\cite{H01,HW01} obtained accurate upper and lower bounds from
the moments equations.

The results, conclusions and criticisms of Bowen et
al\cite{BMFM12} are at variance with all what has been established
after several years of study in the field of non-Hermitian
PT-symmetric Haniltonians\cite{B07}. The purpose of this comment
is to analyse their calculations to verify if such criticisms are
well founded. In section~\ref{sec:DM} we briefly review the
diagonalization method used by the authors, in
section~\ref{sec:examples} we analyse some of the models used by
the authors to draw their conclusions; finally, in
section~\ref{sec:conclusions} we summarize the main results and
draw our own conclusions.

\section{The method}

\label{sec:DM}

Bowen et al\cite{BMFM12} calculated the eigenvalues of the class of
Hamiltonian operators
\begin{equation}
H=p^{2}+sx^{N}  \label{eq:Hamilt}
\end{equation}
for $s=1,-1,i$ and $N=2,3,4,6,8$. They resorted to matrix representations of
the operators in the basis set of eigenfunctions $\{\left| n\right\rangle
,n=0,1,\ldots \}$ of the harmonic oscillator ($s=1$, $N=2$) and diagonalized
truncated Hamiltonian matrices $\mathbf{H}^{(M)}\mathbf{=}\left(
H_{mn}\right) _{m,n=0}^{M-1}$, where $H_{mn}=\left\langle m\right| H\left|
n\right\rangle $, for each of those cases. If the eigenvalues of the
truncated matrices converge as $M$ increases then the limits of those
sequences approach the eigenvalues of the operator (\ref{eq:Hamilt}).

The characteristic polynomial for the matrix $\mathbf{H}^{(M)}$ will exhibit
$M$ roots $W_{n}^{(M)}$, $n=0,1,\ldots ,M-1$. In the case of Hermitian
operators (for example, $s=1$ and $N$ even) all those roots are real because
the matrix is Hermitian. It is not the case for the non-Hermitian operators.
Furthermore, in the case of the Hermitian operators we know that the
eigenvalues of the matrix approach the eigenvalues $E_{n}$ of the operator
from above $W_{n}^{(M)}>W_{n}^{(M+1)}>E_{n}$. On the other hand, there is no
such variational principle in the case of non-Hermitian operators.
Obviously, one has to be very careful when applying the diagonalization
method (DM from now on) to non-Hermitian operators. In what follows we
discuss some examples.

\section{Examples}

\label{sec:examples}

\subsection{Case $N=2$}

When $s=1$ the matrix $\mathbf{H}$ is diagonal and yields the eigenvalues of
the harmonic oscillator exactly: $E_{n}=2n+1$, $n=0,1,\ldots $. On the other
hand, for $s=-1$ the eigenvalues of the matrix do not converge as $M$
increases. Therefore, they are meaningless and bear no relation with the
eigenvalues of the Hamiltonian $H=p^{2}-x^{2}$. Surprisingly, the authors
argued that ``This Hamiltonian has a spectrum with odd symmetry about zero
energy''. If one solves the eigenvalue equation with the appropriate
boundary conditions in the complex-$x$ plane one obtains purely imaginary
eigenvalues: $E_{n}=\pm (2n+1)i$.

The authors went even further and stated that ``Bender has also
asserted that the spectrum of the simple harmonic oscillator (SHO)
($r=0$) with a negative force constant has a discrete negative
spectrum that is the negative of the positive force constant SHO;
that is, $E_{n}=-\hbar \omega (n+1/2)$''. However,
Bender\cite{B07} never draw such wrong conclusion. He discussed
the harmonic oscillator $H=p^{2}+\omega ^{2}x^{2}$ with
eigenvalues $E_{n}(\omega )=(2n+1)\omega $ and eigenfunctions
$\psi _{n}(\omega ,x)$. The eigenfunctions behave asymptotically
as $\psi _{n}(\omega ,x)\sim e^{-\omega x^{2}/2}$ when
$|x|\rightarrow \infty $. If we substitute $-\omega $ for $\omega
$ the eigenvalues change sign but the eigenfunctions are no longer
square integrable.\ However, if we rotate the variable $\pi /2$
counterclockwise then the resulting eigenfunction $\varphi
_{n}(\omega ,q)=\psi _{n}(-\omega ,iq)$ is square integrable. It
is quite obvious that the substitution of $-\omega $ for $\omega$
changes the sign of the eigenvalues but the force constant
($\propto \omega ^{2}$) does not change. In fact,
Bender\cite{B07} states that ``Notice that under the rotation that replaces $%
\omega $ by $-\omega $ the Hamiltonian remains invariant, and yet the signs
of the eigenvalues are reversed!'' Therefore, it seems that Bowen et al\cite
{BMFM12} misread Bender's argument.

\subsection{Case $N=3$}

When $s=1$ the eigenvalues of the truncated matrices do not converge as $M$
increases. However, the authors state that ``Here the spectrum was almost
symmetric about zero...'' in spite of the fact that the roots of the secular
determinants are not valid approximations to the eigenvalues of the
differential operator.

The only interesting case is undoubtedly the PT-symmetric
Hamiltonian operator for $s=i$. According to the authors ``The
calculation of the spectrum for the potential $V=ix^{3}$ yielded a
complex spectrum''. In this case the wedges in the complex-$x$
plane where $\psi (x)$ vanishes exponentially as $|x|\rightarrow
\infty $ contain the real axis\cite{BB98}. Therefore, one expects
the DM to yield meaningful results. Our calculation shows that the
complex eigenvalues of the truncated matrices do not converge as
$M$ increases, but there are real ones that certainly converge
towards the results obtained by Bender and Boettcher\cite{BB98} by
means of the WKB method and numerical integration. In fact, Bender
and Boettcher\cite {BB98,BB97} discussed the calculation of the
eigenvalues by means of the DM. They concluded that the method is
only useful when $1<N<4$ and that the convergence to the exact
eigenvalues is slow and not monotone because the Hamiltonians are
not Hermitian. Table~\ref{Table:DM} shows the convergence of the
lowest eigenvalues of the truncated matrices towards those
obtained by means of the Runge-Kutta method (RK) and the WKB
approach; they are real and positive as argued by Bender and
Boettcher\cite{BB97}.

An interesting feature of the DM for non-Hermitian operators is
that the characteristic polynomial of degree $M$ does not exhibit
$M$ real roots as in the case of the Hermitian matrices. In the
present case the truncated matrices also exhibit many complex
eigenvalues but they do not converge as $M $ increases. Another
interesting feature is the behaviour of the approximate
eigenvalues with respect to a scaling factor. Instead of using the
eigenfunctions $\psi _{n}(x)=\left\langle x\right| \left.
n\right\rangle $ of the harmonic oscillator $p^{2}+x^{2}$ we can
try an alternative calculation with the scaled eigenfunctions
$\alpha ^{1/2}\psi _{n}(\alpha x)$, where $\alpha $ is
an adjustable scaling factor. In the case of Hermitian operators $%
W_{n}^{(M)}(\alpha )$ exhibits a minimum because of the variational
principle. On the other hand, in the case of the complex potential $V=ix^{3}$
the approximate eigenvalue oscillates and exhibits a kind of plateau with
oscillations of smaller amplitude. The optimal value of $\alpha $ is
somewhere in this region. For example, we find that $\alpha \approx 1.4$ is
more convenient than the scaling parameter $\alpha =1$ used in the
calculation shown in Table~\ref{Table:DM}. However, it is our purpose to
show here only the results for the same basis set chosen by Bowen et al\cite
{BMFM12}.

\subsection{Case $N\geq 4$}

In these cases the DM only yields meaningful results for $s=1$ and $N$ even;
that is to say: for the trivial Hermitian operators. When $s=-1$ or $s=i$
the eigenvalues of the truncated Hamiltonian matrices do not converge and,
consequently, they are not eigenvalues of the Hamiltonian operator. The
failure of the naive application of the DM to the PT-symmetric cases is not
surprising because the wedges in the complex-$x$ plane where $\psi (x)$
vanishes exponentially as $|x|\rightarrow \infty $ do not contain the real
axis\cite{BB98}.

The PT-symmetric Hamiltonian operator $H=p^{2}-x^{4}$ deserves special
attention because several authors have proved that it is isospectral to the
Hermitian one $H=p^{2}+4x^{4}-2x$\cite{B07} (and references therein).
Apparently, Bowen et al\cite{BMFM12} were not aware of this relationship
that could have convinced them that the former Hamiltonian does already have
a positive spectrum.

\section{Conclusion}

\label{sec:conclusions}

It is clear that the discrepancy between the results of Bowen et
al\cite {BMFM12} and Bender and Boettcher\cite{BB98} is merely due
to the fact that the DM used by the former authors does not apply
to some of the problems studied. Their conclusions were based on
eigenvalues of the Hamiltonian matrices that do not converge. They
only obtained meaningful results for the trivial cases of
Hermitian Hamiltonians given by $s=1$ and $N$ even. In the only
other selected case where the DM is expected to yield reasonable
results, namely $V(x)=ix^{3}$, the authors failed to find the
converging real positive roots and simply focussed on the complex
ones that do not converge.

\begin{table}[tbp]
\caption{First eigenvalues of the truncated matrices of dimension $M$ for $%
V(x)=ix^3$}
\label{Table:DM}
\begin{center}
\begin{tabular}{lllll}
$M$ & \multicolumn{1}{c}{$E_0$} & \multicolumn{1}{c}{$E_1$} &
\multicolumn{1}{c}{$E_2$} & \multicolumn{1}{c}{$E_3$} \\ \hline
10 & 1.156101684 & 3.73083496 & - & - \\
15 & 1.156038818 & 4.14942907 & - & - \\
20 & 1.156383056 & 4.109441589 & - & - \\
25 & 1.156258544 & 4.109537412 & 7.553497517 & - \\
30 & 1.156267013 & 4.109170441 & 7.562399797 & 11.24884001 \\
35 & 1.156266986 & 4.109228991 & 7.562011977 & 11.31452225 \\
40 & 1.156267082 & 4.109228365 & 7.562284307 & 11.31372188 \\
45 & 1.156267072 & 4.109228831 & 7.562273020 & 11.31452360 \\
50 & 1.156267072 & 4.109228753 & 7.562274330 & 11.31442188 \\
55 & - & 4.109228754 & 7.562273854 & 11.31442413 \\
60 & - & 4.109228753 & 7.562273860 & 11.31442176 \\
65 & - & 4.109228753 & 7.562273854 & 11.31442184 \\
70 & - & 4.109228753 & 7.562273855 & 11.31442182 \\
75 & - & - & 7.562273855 & 11.31442182 \\
80 & - & - & - & 11.31442182 \\ \hline
RK & 1.156267072 & 4.109228752 & 7.562273854 & 11.314421818 \\ \hline
WKB & 1.0943 & 4.0895 & 7.5489 & 11.3043 \\ \hline
\end{tabular}
\end{center}
\end{table}

\end{document}